\documentclass[prd,preprint,superscriptaddress,preprintnumbers,eqsecnum,showpacs,nofootinbib,nobibnotes]{revtex4-1}

\usepackage{amsmath}
\usepackage{amssymb}
\usepackage{graphicx}
\usepackage{color}
\usepackage{xspace}
\usepackage{boondox-cal}
\usepackage[utf8]{inputenc}

\definecolor{purple}{rgb}{0.5,0,0.5}
\definecolor{blue}{rgb}{0.0,0,0.9}
\definecolor{prdblue}{rgb}{0.133,0.118,0.498}

\def\s#1{{\scriptscriptstyle #1}}

\def\noeq#1{(\ref{#1})}
\def\1eq#1{Eq.~(\ref{#1})}

\def\2eqs#1#2{Eqs.~(\ref{#1}) and~(\ref{#2})}
\def\3eqs#1#2#3{Eqs.~(\ref{#1}),~(\ref{#2}) and~(\ref{#3})}

\def\ie{{\it i.e.}, }
\def\eg{{\it e.g.}, }

\def\h{\sigma}

\def\tbarc{\bar{\mathcal{c}}^*}
\def\barc{\bar c^*}
\def\tR{\mathcal{R}}
\def\G{\Gamma}
\def\tG{\widetilde\Gamma}

\def\cw{c_\s{W}}
\def\sw{s_\s{W}}
\def\MW{M_\s{W}}
\def\MZ{M_\s{Z}}
\def\X{X}
\def\c#1{\xi_{#1}}

\begin{document}


\title{Off-shell renormalization in Higgs effective field theories}

\author{Daniele Binosi}
\email{binosi@ectstar.eu}
\affiliation{European Centre for Theoretical Studies in Nuclear Physics and Related Areas (ECT$^\ast$) and Fondazione Bruno Kessler\\ Villa Tambosi, Strada delle Tabarelle 286, I-38123 Villazzano (TN), Italy}
\author{Andrea Quadri}
\email{andrea.quadri@mi.infn.it}
\affiliation{\mbox{Dip. di Fisica, Universit\`a degli Studi di Milano and INFN, Sezione di Milano,} via Celoria 16, I-20133 Milano, Italy}

\date{28 September 2017}

\begin{abstract}

The off-shell one-loop renormalization of a Higgs effective field theory possessing a scalar potential $\sim\left(\Phi^\dagger\Phi-\frac{v^2}2\right)^N$ with $N$ arbitrary is presented. This is achieved by renormalizing the  theory once reformulated in terms of two auxiliary fields $\X_{1,2}$, which, due to the invariance under an extended Becchi-Rouet-Stora-Tyutin symmetry, are tightly constrained by functional identities. The latter allow in turn the explicit derivation of the mapping onto the original theory, through which the (divergent) multi-Higgs amplitude are generated in a purely algebraic fashion. We show that, contrary to naive expectations based on the loss of power counting renormalizability, the Higgs field undergoes a linear Standard Model like redefinition, and evaluate the renormalization of the complete set of Higgs self-coupling in the $N\to\infty$ case.
\end{abstract}


\maketitle

\section{Introduction}

In this paper we extend the techniques and tools developed in the context of algebraic renormalization of gauge theories~\cite{Piguet:1995er,Ferrari:1999nj,Quadri:2003ui,Quadri:2003pq,Quadri:2005pv,Grassi:1999tp,Grassi:2001zz} in order to study the ultraviolet (UV) behavior of a particular class of Higgs Effective Field Theories (HEFTs) whose higher dimensional operators are constructed out of arbitrary powers of the gauge-invariant combination $\Phi^\dagger \Phi-\frac{v^2}2$, \ie for which the Beyond the Standard Model (BSM) Higgs potential is written as
\begin{align}
	V\sim\sum_{j=3}^\infty g_{2j}\left(\Phi^\dagger \Phi-\frac{v^2}2\right)^j,
	\label{introV}
\end{align}
 with $\Phi$ the standard Higgs SU(2) doublet (see \1eq{Phi} below) and $v$ its vacuum expectation value (vev). 

There are several physical motivations to study such a class of potentials. In fact, 
the shape and structure of the spontaneous symmetry breaking potential
are key ingredients in the program of experimental study of the Higgs boson properties; yet, the scalar couplings are poorly known at the moment. Indeed, a huge effort is currently under way in order  to provide the
best experimental constraints at the LHC on the Higgs self-couplings,
as well as in order to assess their phenomenological impact 
in terms of possible deviations from the SM values (for a recent review see, {\it  e.g.},~\cite{deFlorian:2016spz}).

More generally, different choices of the analytical potential $V$ lead to very different scenarios for electroweak symmetry breaking than the simple SM mechanism~\cite{Arkani-Hamed:2015vfh}.
As a consequence, exploration of the global structure of the Higgs potential is an important question that has to be addressed, beyond the small, quadratic oscillations around the vev to be probed by LHC. For instance, it is well-known that a dimension 6 operator (corresponding to $j=3$ in the above equation) can lead to a first order electroweak phase transition~\cite{Zhang:1992fs,Zhang:1994qb,Grojean:2004xa}.
Neverthless, no reliable statement can be made even at the one-loop order by truncating the series in~\1eq{introV} to a polynomial: quantum corrections induce operators of higher dimension, and, therefore, knowledge of the full renormalization of the analytic potential is required~\cite{DiLuzio:2017tfn}.

Moreover the UV properties and operator mixing of BSM theories of this type, and in particular the ones involving dimension 6 operators, have in fact received substantial attention in recent years, both for phenomenological reasons~\cite{Degrassi:2017ucl,Kribs:2017znd,Deutschmann:2017qum,Degrassi:2016wml,deFlorian:2016spz}, and as a consequence of the discovery of surprising cancellation patterns in the one-loop on-shell anomalous dimensions~\cite{Jenkins:2013zja,Jenkins:2013wua,Alonso:2013hga} which have been traced back to holomorphy~\cite{Cheung:2015aba} and to the remnant of embedding supersymmetry transformations approximately constraining these (non-supersymmetric) HEFTs~\cite{Elias-Miro:2014eia}. 

The novel approach introduced here will allow for the first time, to the best of our knowledge, to achieve the complete off-shell renormalization of the Higgs sector of these BSM theories at the one-loop level. This is different form the standard approach to HEFTs, where one is usually interested in physical $S$-matrix quantities, so that the equations of motion are freely used in order to simplify the basis of operators involved~\cite{Burges:1983zg,Leung:1984ni,Buchmuller:1985jz,Grzadkowski:2003tf,Fox:2007in,AguilarSaavedra:2008zc,AguilarSaavedra:2009mx,Grojean:2006nn,Grzadkowski:2010es,Contino:2013kra}. As a consequence one is limited to the classification of {\em on-shell} UV divergences only. This means in turn that while one can compute the anomalous dimensions $\gamma$ of physical operators, the $\beta$ functions of the corresponding couplings are out of reach, since in order to evaluate the latter one needs to take into account the contribution of wave function renormalizations. On the other hand, evaluation of the latter requires to take properly into account (Higgs) field redefinitions which in a generic HEFT, due to the loss of  power counting renormalizability, can be polynomial and derivative dependent, being subjected only to the requirement of being induced by a canonical transformations with respect to the Batalin-Vilkovisky bracket associated with the classical Becchi-Rouet-Stora-Tyutin (BRST) differential~\cite{Gomis:1995jp}.   

In this respect we will establish the somewhat surprising result that for HEFTs characterized by the potential~\noeq{introV} the Higgs wave-function renormalization is purely SM-like, so that the Higgs field redefinition is the familiar linear one known from the SM power counting renormalizability. This is not obvious, since, in an effective field theory, wave-function renormalizations are in general superseded by a canonical transformation~\cite{Gomis:1995jp}. This implies in turn that for the Higgs field one could have a polynomial field redefinition, {\it e.g.}, $\h \rightarrow Z_\h^{1/2} \h + a_1 \h^2 + a_2 \sigma^3 + \cdots$, so that, under such a transformation, the quadratic part of the action would also affect Green's functions with a higher number of external legs, at variance with the power-counting renormalizable case. This indeed does not happen in the case at hand.

In addition we will be able to write in a closed form the renormalization of the $g_{2j}$ couplings considering the complete tower of operators in~\noeq{introV}. This paves the way to the computation of the  $\beta$ functions of the $g_{2j}$ BSM couplings without resorting to any polynomial approximation of $V$ in the tree-level Lagrangian, that will necessarily give rise to instabilities at one loop order due to the generation of additional divergences not contained in the tree-level action  and thus new operators absent in the truncated classical potential,~see,~\eg~\cite{DiLuzio:2017tfn}.

The paper is organized as follows. In Section~\ref{sec:Xfields} we will introduce our formalism and in particular show how to reformulate the electroweak SSB mechanism using as a dynamical variable the gauge invariant combination $\Phi^\dagger \Phi-\frac{v^2}2$. This is achieved by introducing a new field $X_2$ together with a Lagrange multiplier $X_1$ enforcing on-shell the condition $X_2 = \frac{1}{v} \left ( \Phi^\dagger \Phi - \frac{v^2}{2} \right )$. 

Specializing then to the particular case of a cubic potential in $X_2$, in Section~\ref{sec:cubic} we show how in this new formulation the BSM operators admit an interpretation in terms of certain external sources coupled to a tower of $X_2$-dependent operators with a better UV behaviour than those of the quantized fields $X_1$ and $X_2$ themselves. In addition, we derive a set of functional identities that fully constrain the dependence of the vertex functional on those fields, to all orders in the loop expansion. As a result, we show that it is much easier to analyze the 1-PI amplitudes in the so-called $\X$-theory (being constrained by the hidden symmetries described by the aforementioned functional identities) and then read-off the needed information in the original (target) theory via a suitable mapping, that we explicitly identify. 

We also discuss how power-counting is realized in this formulation, and identify accordingly all divergent amplitudes, from which one can generate all the multi-Higgs divergent amplitudes in the target theory in a purely algebraic fashion through the mapping previously constructed. Finally we show the rather remarkable property that, at the one-loop level, the derivative dependent operators $\partial_\mu(\Phi^\dagger\Phi)\partial^\mu(\Phi^\dagger\Phi)$ and $(\Phi^\dagger D_\mu \Phi)(\Phi^\dagger D_\mu \Phi)$ are not radiatively generated, despite being compatible with both the functional identities and the associated power counting. This can be seen as a consequence of some seemingly accidental cancellations that are transparent once expressed in terms of the external sources amplitudes of the $\X$ theory. This fact has phenomenological relevance, since if such operators were radiatively generated, their mixing with $\left(\Phi^\dagger \Phi-\frac{v^2}2\right)^3$ would occur, ultimately invalidating analyses where they are excluded in the evaluation of physical observables (as done, {\it e.g.}, in~\cite{Degrassi:2016wml}). 

Next, in Section~\ref{sec:arb} we show that the cubic case is actually much more than an informative warm-up exercise, and proceed to describe how we can generalize the procedure developed in that case to a potential involving arbitrary powers of the operator $\left(\Phi^\dagger \Phi-\frac{v^2}2\right)$. Since the power counting does not change in the $X$-theory, the divergent amplitudes are the same as the ones identified in the cubic case, and all divergent multi-Higgs amplitudes in the target theory can be again generated in an exclusively algebraic way through a generalized mapping that we explicitly work out. 

Section~\ref{sec:renorm} describes finally how one can carry out the full off-shell one-loop renormalization of the model. We first renormalize the $\X$-theory, then show that the vev and Higgs wavefunction renormalizationn are the SM ones, therefore excluding non-linear field redefinitions in the Higgs sector, and, finally, renormalize the target theory determining in particular the renormalization of the BSM couplings $g_{2j}$. Our conclusions are then presented in Section~\ref{sec:concl}. The paper ends with two Appendices. In Appendix~A we report the Higgs one, two and three-point function in the target theory for the cubic case. In Appendix B we provide instead the UV divergent parts of the divergent 1-PI amplitudes in the $\X$-theory for a cubic potential, and construct explicitly the multi-Higgs divergent amplitudes in the target theory.

\section{\label{sec:Xfields}$\X$ fields}

As has been recently shown in~\cite{Quadri:2016wwl}, it is possible to reformulate the electroweak SSB mechanism using as a dynamical variable the gauge invariant combination $\Phi^\dagger \Phi-\frac{v^2}2$, where $\Phi$ is the standard scalar field SU(2) doublet, \ie
\begin{align}
	\Phi=\frac{1}{\sqrt{2}}\left(\begin{array}{c}i\phi_1+\phi_2\\ \phi_0+i\phi_3\end{array}\right),
	\label{Phi}
\end{align}
with $\phi_i$ the Goldstone's bosons, $\phi_0$ the would be Higgs field and $v$ its expectation value, $\langle \phi_0\rangle=v$. It is the purpose of this paper to illustrate why it is advantageous to do so; in what follows we rather recall how this reformulation can be technically accomplished. 

To this end, let us split the tree-level action of our theory in three parts, writing
\begin{align}
	\Gamma=\Gamma_\s{\mathrm{SM}}+\Gamma_{\tiny{\mbox{SSB}}}+\Gamma_{\tiny{\mbox{SRC}}}. 
\end{align}
Above, $\Gamma_\s{\mathrm{SM}}$ represents the usual SM action comprising the Yang-Mills, fermion, Yukawa, (linear) gauge fixing and ghost terms; $\Gamma_{\tiny{\mbox{SSB}}}$ replaces instead the SSB Higgs term and reads (omitting, here and in the following, space-time arguments from all fields whenever no confusion can arise) 
\begin{align}
    \Gamma_{\tiny{\mbox{SSB}}}&=\int
    \left[D_\mu\Phi^\dagger D^\mu\Phi-\frac{M^2-m^2}2 X_2^2
    -\frac{m^2}{2v^2} \Big ( \Phi^\dagger\Phi-\frac{v^2}2 \Big )^2-\overline{c}\left(\square+m^2\right) c
    \right.\nonumber \\
    &
    +\left.\frac1v\left(X_1+X_2\right)(\square+m^2)\left(\Phi^\dagger\Phi-\frac{v^2}2-vX_2
    \right)+V(X_2)\right],
    \label{action}
\end{align}
with $D$ the usual covariant derivative and $V(X_2)$ a generic potential in the $X_2$ field\footnote{With respect to the conventions of~\cite{Quadri:2016wwl} we have reinstated the dependence on $m^2$ (which is allowed in the most general power-counting renormalizable theory in the $\X$-formalism~\cite{Quadri:2006hr}) and replaced $M^2 \rightarrow M^2 - m^2$ in such a way that $M$ is here the mass of the physical scalar resonance.}. The equation above makes it clear that only $X_2$ is a genuine field; $X_1$ is instead a Lagrange multiplier, whose equation of motion $\Gamma_{X_1}=0$ enforce the condition 
\begin{align}
	X_2= \frac{1}{v} \left( \Phi^\dagger \Phi - \frac{v^2}{2} \right).
\end{align}

The advantage of this reformulation is that the scalar Higgs field can be seen as a quantum fluctuation around the SU(2) constraint $\Phi^\dagger \Phi - \frac{v^2}{2}=0$, {\it i.e.}, the sphere in the field space configuration spanned by the field coordinates $(\h,\phi_a)$. If such a fluctuation is frozen out, one recovers the St\"uckelberg model~\cite{Quadri:2006hr}; on the other hand, if the non-linear constraint is relaxed by a non-vanishing quantum fluctuation $X_2$, one gets back a model that has the same physical particle content as the usual Higgs theory.

In fact, the field $X_2$ describes the physical scalar excitation, whereas $\sigma=\phi_0-v$ can be traded in favour of the unphysical mass eigenstate combination $\sigma'=\sigma-X_1-X_2$. Both $\sigma'$ and $X_1$ have mass $m^2$ and their propagator differ by a sign. They cancel against each other in all physical amplitudes\footnote{A rigorous all-orders proof of this result has been given in~\cite{Quadri:2016wwl,Quadri:2006hr}.}. Accordingly the mass parameter $m$ is an unphysical parameter that {\it must drop out of all physical amplitudes}, leaving only the dependence on the physical Higgs mass $M$. At tree-level this is easily understood, since by going on-shell with $X_1$ and replacing $X_2$ with $\Phi^\dagger \Phi - \frac{v^2}{2}$ the $m^2$-dependent terms in the first line of Eq.(\ref{action}) cancels out, and one is left with the conventional quartic Higgs potential $V_{\tiny{\mbox{SM}}}=\frac{M^2}{2v^2} \left(\Phi^\dagger \Phi - \frac{v^2}{2} \right)^2$. At higher orders, the $m^2$-independence of the results gives a powerful way to check calculations performed in the $\X$-theory.  

In this framework the SM corresponds to the simplest approximation, namely the quadratic action for the scalar (SU(2) singlet) $X_2$. One can then add either self-interactions in $X_2$ (giving rise to a BSM potential, which we will study in detail in the present paper) or more general gauge-invariant higher dimensional  couplings with the SM fermions and gauge bosons. One might even think of a more complicated dynamics, {\it e.g.}, in the relaxion models~\cite{Graham:2015cka,Espinosa:2015eda,Hook:2016mqo} where the Higgs mass squared is promoted to a function of a slowly rolling field $\varphi$ (the axion being the simplest possibility) that scans the Higgs mass and stabilizes the latter since the potential barrier increases as a consequence of the increase of the Higgs vacuum expectation  value, ultimately preventing $\varphi$ from rolling down further. When addressing the quantum stability for this class of models, a series of operators involving  the singlet fluctuation $X_2$ and the field  $\varphi$ naturally arise, and the formalism developed here is expected to provide a useful tool in order to assess the renormalization properties of such models. In particular, anticipating some technical aspects that will be  clarified soon, we expect to be possible to study the UV properties of composite operators of this type by generalizing the results presented in this paper to  $\varphi$-dependent external sources coupled to powers of $X_2$.

Finally, one has
\begin{align}
	\Gamma_{\tiny{\mbox{SRC}}}=\int\overline{c}^*\left(\Phi^\dagger\Phi-\frac{v^2}2-vX_2\right)+\Gamma_{\tiny{V}}.
\end{align}
Here $\barc$ is a source (usually referred to as an {\it antifield}) coupled to the on-shell null operator $\Phi^\dagger \Phi - \frac{v^2}{2} -vX_2$ which is needed to implement the $X_1$ equation of motion; $\Gamma_{\tiny{V}}$ collects instead all the additional sources needed for implementing the $X_2$ equation of motion, whose number depends on the actual form of the $X_2$ potential, as we shall soon see.

The action $\Gamma$ is  invariant under the extended Becchi-Rouet-Stora-Tyutin (BRST) operator $s+\mathcal{s}$ where $s$ is the standard BRST operator associated to the SU(2)$\times$U(1) gauge invariance, while $\mathcal{s}$ implements algebraically the (SU(2)-invariant) constraint, namely
\begin{align}
    \mathcal{s}X_1&=vc;& \mathcal{s}\Phi&=\mathcal{s}X_2=\mathcal{s}c=\mathcal{s}\overline{c}^*=0;&\mathcal{s}\overline{c}&=\Phi^\dagger\Phi-\frac{v^2}{2}-vX_2.
\end{align}
Notice that the two BRST operators anticommute, $\{s,\mathcal{s}\}=0$, and they are both nilpotent, $s^2=\mathcal{s}^2=0$.

When going on-shell with both $X_1$ and $X_2$ (that is imposing the equations of motion $\Gamma_{X_i}=0$, $i=1,2$) the 1-PI amplitudes generated will coincide with the ones obtained in the standard formalism. We call the resulting theory the target theory; in particular for $V(X_2)=0$ one recovers that the target theory coincides with the SM. Clearly, then, the potential $V(X_2)$ contains all the BSM operators expressed as polynomials in the $X_2$ field itself. For example, looking at dimension 6 operators, $V(X_2)$ might contain terms like $X_2^3$, and/or ordinary derivatives, \eg the canonically normalized $X_2$ kinetic term $\frac12\partial_\mu X_2\partial^\mu X_2$; on-shell these terms would  map onto the corresponding operators $\frac{1}{v^3} \left(\Phi^\dagger\Phi-\frac v2\right)^3$ and $\frac{1}{v^2} \partial_\mu(\Phi^\dagger\Phi)\partial^\mu(\Phi^\dagger\Phi)$ respectively.

However, in the $\X$ formulation some BSM operators admit a reformulation in terms of a suitable set of external sources coupled to a tower of $X_2$-dependent operators with a better UV behaviour than those of the quantized fields $X_1$ and $X_2$. In addition, a  set of functional identities exists fully constraining the dependence of the vertex functional on those fields, to all orders in the loop expansion. As a result, it is much easier to analyze the 1-PI amplitudes in the $\X$ theory (being constrained by the hidden symmetries described by the aforementioned functional identities) and then read-off the needed information in the target theory via a suitable mapping onto it (that we need to identify).

\section{\label{sec:cubic}The cubic potential case~$X_2^3$}

As a simple yet illustrative warm-up exercise, we will now study in detail the BSM theory obtained when choosing the potential 
\begin{align}
	V(X_2)=g_6 \Lambda X_2^3,
	\label{cubic}
\end{align}
where $\Lambda$ is a mass parameter introduced in order to make the coupling constant $g_6$ dimensionless. This is the lowest dimension BSM operator with no derivative that can be built in the $\X$ theory; yet, in the target theory, it plays an important role in the phenomenological study of the Higgs potential at the LHC~\cite{deFlorian:2016spz}. Thus, after constructing explicitly the mapping between the $\X$ and the target theory, we will then:
\begin{itemize}
    \item[i)] Reconstruct the set of one-loop counterterms required to renormalize the theory at one loop level in the sector involving only the Higgs field $\h$, starting from the (few) UV divergent amplitudes involving the external sources of the $\X$-theory;
    \item[ii)] Show that at the one-loop level the operator $\partial_\mu(\Phi^\dagger\Phi)\partial^\mu(\Phi^\dagger\Phi)$ is not radiatively generated (despite being compatible with both the functional identities and the associated power counting), as a consequence of some accidental cancellations that are transparent once expressed in terms of the external sources amplitudes of the $\X$ theory. This is of phenomenological relevance since if such an operator were radiatively generated, the mixing between $(\Phi^\dagger \Phi)^3$ and $\partial_\mu(\Phi^\dagger\Phi)\partial^\mu(\Phi^\dagger\Phi)$ would occur and one could not exclude the latter operator in the evaluation of physical observables at this order of approximation, as has been done, {\it e.g.}, in~\cite{Degrassi:2016wml}.
\end{itemize}
The results obtained will go beyond a mere illustrative purpose, and, in fact, pave the way to the complete one-loop analysis and (algebraic) renromalization of a generic potential containing arbitrary powers in the $X_2$ field (but no derivative terms) that will be carried out in Sect.~\ref{sec:renorm}.

\subsection{\label{map}$\X$ functional identities and mapping to the target theory}

In the presence of the potential~\noeq{cubic}, one then needs to introduce in the action~\noeq{action} a single additional source $R$ coupled to the field $X_2^2$:
\begin{align}
	\Gamma_{\tiny{V}}=\int RX_2^2
\end{align}
 The reason is simple: the derivative of the action with respect to $X_2$ in the presence of the trilinear interaction vertex~\noeq{cubic} gives rise to the composite operator $X_2^2$, which, being non-linear in the quantized fields, needs to be defined through the coupling to a suitable external source~\cite{Piguet:1995er}. Then the equation of motion of the $X_1$ and $X_2$ fields read respectively\footnote{We will denote by a subscript the functional derivatives with respect to the corresponding field; thus for a generic field $\varphi$ we have
\begin{align}
 \Gamma_{\varphi_x}&=\frac{\delta\G}{\delta \varphi(x)};& \G_{\varphi_x\varphi_y}&=\frac{\delta^2\G}{\delta \varphi(x)\delta\varphi(y)}, \nonumber
\end{align}
and so on. As before, whenever no confusion can arise, space-time arguments will be omitted and we will simply write $\G_\varphi$, $\Gamma_{\varphi\varphi}$, etc. Finally, when considering 1-PI functions, evaluation at zero external sources and fields after the functional differentiation is assumed.}
\begin{align}
    \G_{X_1}& =\frac{1}{v} \left(\square + m^2\right) 
    \G_{\barc}, \label{X1eom} \\
    \G_{X_2}& = \frac{1}{v} \left(\square + m^2\right) 
    \G_{\barc}
    + 3 g_6 \Lambda 
    \G_{R}-  \left(\square + m^2\right)X_1\nonumber \\
     &  -  \left(\square + M^2\right)X_2 + 2 R X_2- v \barc .
\label{X2eom}
\end{align}
These equations, which are valid to all orders in the perturbative expansion, imply that (starting at the one-loop level) the whole dependence of the vertex functional $\G$ on $X_1,X_2$ enters only through the combinations
\begin{align}
	\tR &= R + 3 g_6 \Lambda X_2;&  \tbarc &= \barc + \frac{1}{v}\left(\square + m^2\right)\left(X_1+X_2\right).
\label{ext.srcs}
\end{align}
In fact, due to the differentiation chain rule, if we write $\Gamma=\Gamma[\tR,\tbarc]$ one has (omitting the spacetime integration symbol) 
\begin{align}
\Gamma_{X_2}\supset\Gamma_R \frac{\delta R}{\delta\tR}\frac{\delta\tR}{\delta{X_2}}&=3g_6\Lambda\Gamma_R;& 
\Gamma_{X_1}\supset\Gamma_{\bar c^*}\frac{\delta\bar c^*}{\delta\tbarc}\frac{\delta\tbarc}{\delta X_1}=\frac{1}{v} \left(\square + m^2\right)\Gamma_{\bar c^*}.
\end{align}
Hence one can trade amplitudes involving external $X_1$ and/or $X_2$ legs for amplitudes with insertions of the external sources $\tbarc$ and/or $\tR$.

From these Green's functions we can recover the corresponding ones in the target theory by going on-shell with the $X_1$ and $X_2$ fields. If one is interested only in a lowest order perturbative analysis, it is sufficient to consider the tree-level equations of motion of those fields and substitute them back into the right-hand side of~\1eq{ext.srcs}. The $X_1$-equation yields then the constraint
\begin{align}
    X_2 = \frac{1}{v} \left(\Phi^\dagger \Phi - \frac{v^2}{2} \right) = \h + \frac{1}{2} \frac{\h^2}{v} + \frac{1}{2} \frac{\phi_a^2}{v}, 
\label{sol.X1eom}
\end{align}
whereas the $X_2$ field equation of motion reduces to
(at zero external sources $R=0$, $\barc=0$)
\begin{align}
    \left(\square + m^2\right) \left(X_1 + X_2\right) = - \left(M^2 - m^2\right) X_2 + 3 g_6 \Lambda X_2^2.
\label{sol.X2eom}
\end{align}
Substituting then \2eqs{sol.X1eom}{sol.X2eom} into \1eq{ext.srcs} one obtains the new sources\footnote{Notice that despite the non-local nature of the solution to the $X_2$-equation of motion (due to the presence of the Klein-Gordon operator) no poles arise in the vertex functional of the target theory after going on-shell with $X_1$ and $X_2$, as a consequence of~\1eq{X2eom}.}
\begin{align}
    \tR &\to 3 g_6 \frac\Lambda v \left( \Phi^\dagger \Phi - \frac{v^2}{2} \right),\nonumber \\
    \tbarc &\to -\frac{1}{v^2}\left(M^2 - m^2\right) \left( \Phi^\dagger \Phi - \frac{v^2}{2} \right) +\frac{3 g_6\Lambda}{v^3}  \left( \Phi^\dagger \Phi - \frac{v^2}{2} \right)^2,
\label{repl.rule}
\end{align}
which constitute the sought for scalar sector $\X$-theory mapping onto the target theory at the one-loop level. 

Notice finally that the replacement rules 
in the $X$-theory in Eq.(\ref{ext.srcs})
above are valid to all orders in the loop expansion. Yet at orders higher than one the mapping to the target theory changes since loop corrections to the $\X$-equations of motion have to be considered when substituting the on-shell solution for $\X$, which in general will not be any more given by their tree-level approximation of~\2eqs{sol.X1eom}{sol.X2eom}.

\subsection{One-loop Higgs one- and two-point functions in the target theory}

To see how the mapping works in practice, we show how to reconstruct 
the Higgs one- and two-point function to lowest order in the target theory
from the $\X$ 1-PI functions.

In the one-point sector one has two functions to consider: $\G^{(1)}_R$ and $\G^{(1)}_{\barc}$. Then,~\2eqs{sol.X1eom}{repl.rule} show that their contribution to the Higgs one-point function in the target theory is
\begin{align}
    \int\!\G^{(1)}_{R_x}R_x&\to\int\!\G^{(1)}_{R_x}\tR_x\underset{\h\ \mathrm{term}}{\to}3g_6\Lambda\int\!\G^{(1)}_{R_x}\h_{\!\!x}, \nonumber\\
    \int\!\G^{(1)}_{\barc_x}\barc_x&\to\int\!\G^{(1)}_{\barc_x}\tbarc_x\underset{\h\ \mathrm{term}}{\to}-\frac1{v}\left(M^2-m^2\right)\int\!\G^{(1)}_{\barc_x}\h_{\!\!x},
\end{align}
and, consequently,
\begin{align}
    \widetilde{\G}^{(1)}_{\h}
    =3g_6\Lambda\G^{(1)}_{R}-\frac1{v}\left(M^2-m^2\right)\G^{(1)}_{\barc}.
    \label{1pt}
\end{align}

The two-point sector is slightly more involved, as one needs to consider the $\X$ 1-PI functions with up to two external~$\h,\barc$ or~$R$ legs. To begin with observe that tadpoles in the $\X$ theory will contribute to the target-theory two-point functions as~\2eqs{sol.X1eom}{repl.rule} imply 
\begin{align}
    \int\!\G^{(1)}_{R_x}R_x&
    \underset{\sigma^2\ \mathrm{term}}{\to}\frac{3\Lambda}{2v}g_6\int\!\G^{(1)}_{R_x}\h_{\!\!x}\h_{\!\!x}, \\
    \int\!\G^{(1)}_{\barc_x}\barc_x&
    \underset{\sigma^2\ \mathrm{term}}{\to}-\frac1{2v^2}(M^2-m^2-6g_6\Lambda v)\int\!\G^{(1)}_{\barc_x}\h_{\!\!x}\h_{\!\!x}.\nonumber
\end{align}
Similarly, the $\X$ two-point sector yield the contributions 
\begin{align}
    \int\hspace{-.2cm}\int\!\frac12\G^{(1)}_{R_xR_y}R_xR_y&\underset{\sigma^2\ \mathrm{term}}{\to}
    \frac92g^2_6\Lambda^2\!\int\hspace{-.2cm}\int\!\G^{(1)}_{R_xR_y}\h_{\!\!x}\h_{\!\!y},\nonumber \\
    \int\hspace{-.2cm}\int\!\G^{(1)}_{R_x\h_{\!\!y}}R_x\h_{\!\!y}&\underset{\sigma^2\ \mathrm{term}}{\to}
    3g_6\Lambda\!\int\hspace{-.2cm}\int\!\G^{(1)}_{R_x\h_{\!\!y}}\h_{\!\!x}\h_{\!\!y},\nonumber \\
    \int\hspace{-.2cm}\int\!\frac12\G^{(1)}_{\barc_x\barc_y}\barc_x\barc_y&\underset{\sigma^2\ \mathrm{term}}{\to}
    \frac1{2v^2}(M^2-m^2)^2\int\hspace{-.2cm}\int\!\G^{(1)}_{\barc_x\barc_y}\h_{\!\!x}\h_{\!\!y},\nonumber \\
    \int\hspace{-.2cm}\int\!\G^{(1)}_{\barc_x\h_{\!\!y}}\barc_x\h_{\!\!y}&\underset{\sigma^2\ \mathrm{term}}{\to}
    -\frac1v(M^2-m^2)\int\hspace{-.2cm}\int\!\G^{(1)}_{\barc_x\h_{\!\!y}}\h_{\!\!x}\h_{\!\!y},\nonumber \\
    \int\hspace{-.2cm}\int\!\G^{(1)}_{R_x\barc_y}R_x\barc_y&\underset{\sigma^2\ \mathrm{term}}{\to}-3g_6\frac{\Lambda}{v}(M^2-m^2)\int\hspace{-.2cm}\int\!\G^{(1)}_{R_x\barc_y}\h_{\!\!x}\h_{\!\!y}.
\end{align} 
Summing up all the above BSM contributions to the SM one  $\Gamma^{(1)}_{\h\h}$ yields then the final target-theory function
\begin{align}
    \tG^{(1)}_{\h\h}&=\Gamma^{(1)}_{\h\h}+3g_6\left(\frac{\Lambda}{v} \G^{(1)}_R+2\frac{\Lambda}v\G^{(1)}_{\barc}+2\Lambda\G^{(1)}_{R\h}+3g_6\Lambda^2\G^{(1)}_{RR}\right)\nonumber \\
    &-\frac1{v^2}\left(M^2-m^2\right)\left[\G^{(1)}_{\barc}+2v\G^{(1)}_{\barc\h}+6g_6\Lambda v\G^{(1)}_{R\barc}
    -\left(M^2-m^2\right)\G^{(1)}_{\barc\barc}\right].
    \label{2pt}
\end{align}
 
Notice that at one loop level there is no dependence on $g_6$ arising from the $\X$-theory amplitudes, since at this level the interaction vertex $\sim g_6 X_2^3$ can only contribute to 1-PI Green's functions with at least one external $X_2$-leg. Therefore there is no further $g_6$-dependence in $\tG^{(1)}_{\h\h}$ in addition to the one displayed in~\1eq{2pt}.

Explicit results for the above Green's functions as well as the three-point function $\widetilde{\Gamma}_{\h\h\h}^{(1)}$ are collected in Appendix~\ref{app:3sigma}. In particular, we notice here that the UV divergent parts of all terms appearing on the right-hand side of~\1eq{2pt}  (except the first one) are momentum-independent; we then conclude that {\it the wave function renormalization of the $\h$ field is the same as the SM one.}

This result shows that the linear field redefinition associated with the
wave-function renormalization does not receive contributions depending on 
the BSM coupling $g_6$. In particular, such a field redefinition does not 
contain differential operators (which might in principle appear in 
an effective field theory). Furthermore, we will prove shortly that 
the SM wave-function renormalization is enough to remove all the 
one-loop divergences of the model, together with the redefinition 
of the external sources $R$ and $\barc$ solving the $X$-equations of motion.
Plainly, this implies that there are no polynomial contributions to the 
Higgs field redefinition. This is a non-trivial property since in an
effective gauge field theory one would in general expect the appearance of polynomial, possibly derivative-dependent field redefinitions induced by canonical transformations 
w.r.t. the Batalin-Vilkovisky bracket enconding at the quantum level
the BRST invariance of the classical action \cite{Gomis:1995jp}.
 
\subsection{\label{powerc}Power-counting}

Let us now consider if and how power counting is realized in the $\X$-theory. Let us define $X=X_1+X_2$, and consider first the case in which $g_6=0$. Then, contractions inside loops involving the derivative interaction vertices generated in~\1eq{action} by the term $X \square \Phi^\dagger \Phi$ will always involve either the propagator $\Delta_{XX}$ or $\Delta_{X\h}$. Both propagators however behave like $1/p^4$ for large momenta~\cite{Quadri:2016wwl}, thus ensuring power-counting renormalizability. On the other hand, when $g_6$ is switched on, contractions involving the propagators $\Delta_{X_2 X}$ and $\Delta_{X_2 \h}$ also arise inside loops; as these propagators behave as $1/p^2$ for large momenta, they cannot anymore compensate the additional momentum dependence from the derivative interaction vertices, and thus power-counting renormalizability is violated, as expected.

Nevertheless at one-loop level power-counting rules are very simple also when $g_6 \neq 0$. In fact, recall that we do not need to consider amplitudes involving external $X_1$ and $X_2$ legs as the latter are uniquely fixed by \2eqs{X1eom}{X2eom} once amplitudes involving the external sources $\tbarc$ and $\tR$ are known. Whence, to  lowest order in the perturbative expansion the trilinear interaction vertex $g_6 \Lambda X_2^3$ does not play any role, and the unique source of non-power-counting renormalizability is represented by the interaction vertex $R X_2^2$.

We can limit ourselves to the sector where $X_1=X_2=0$, since the whole dependence on $\X$ will be given by the replacement in~\1eq{ext.srcs}. Then, at one-loop level:
\begin{enumerate}
    
    \item In the $R$-independent sector the model is power-counting renormalizable, and the source $\barc$ has UV dimension $2$, whereas $\h$ has UV dimension $1$. Thus if we limit ourselves to UV-divergent amplitudes only involving $\h$ and $\barc$, we find by power-counting the following eight UV-divergent Green's functions: $\G^{(1)}_{\h},\ \G^{(1)}_{\h\h},\ \G^{(1)}_{\h\h\h},\ \G^{(1)}_{\h^4}$ and $\G^{(1)}_{\barc},\ \G^{(1)}_{\barc\h},\ \G^{(1)}_{\barc\h\h},\  \G^{(1)}_{\barc\barc}$.
    
    \item For one-loop amplitudes involving $R$ and/or $\barc$ only, $R$ behaves as a source with UV dimension $2$; thus by power-counting there are three UV-divergent amplitudes: $\G^{(1)}_R,\ \G^{(1)}_{R\barc},\ \G^{(1)}_{RR}$.
    
    \item When amplitudes also involving external $\h$ legs are considered, one cannot assign UV degree $2$ to $R$ and $1$ to $\h$, since in this case one would
    expect to find UV-divergent amplitudes with one $R$ and up to two $\h$-insertions. However, this is not the case, since amplitudes with the insertion of one $R$ and up to four $\h$-legs are also divergent, as a consequence of the contractions involving the legs of the derivative interaction vertex with the $X_2$-propagator (as explained in detail below).  Classification of UV divergent amplitudes is then carried out as follows: 
    
    \begin{enumerate}
        
        \item The highest UV degree of a diagram involving one or more $R$ sources and a given number of external $\h$-legs is obtained by maximizing the number of internal propagators $\Delta_{\h\h}$ (which drop off as $1/p^2$ for large momenta) and the number of derivative interaction vertices. Consequently we find that $\G^{(1)}_{R\h}$ and $\G^{(1)}_{R\h\h}$ have UV degree $2$ and $\G^{(1)}_{R\h\h\h}$ and $\G^{(1)}_{R\h^4}$ UV degree zero; all remaining amplitudes with one $R$-insertion and a number of $\h$-legs higher than 4 are UV convergent. 
        
        \item Similarly, in the sector with two $R$-external sources there are four logarithmically divergent amplitudes, namely $\G^{(1)}_{RR\h}$, $\G^{(1)}_{RR\h\h}$, $\G^{(1)}_{RR\h\h\h}$ and $\G^{(1)}_{RR\h^4}$. 
        
        \item Finally, in the mixed $R$-$\barc$-$\h$ sector, there are two further logarithmically divergent amplitudes: $\G^{(1)}_{R\barc \h}$ and $\G^{(1)}_{R \barc \h \h}$.
    \end{enumerate}
\end{enumerate}
We stress once again that no such one-loop amplitude will contain a $g_6$-dependent contribution; the whole dependence on $g_6$ arises from the replacement rule in \1eq{repl.rule}. However, this property will not generalize to higher orders, since in that case the trilinear interaction vertex in $X_2$ can appear inside loops. The UV divergent parts of the above amplitudes are reported in \ref{app:cts}.

We notice that by power-counting the Green's function $\G^{(1)}_{R\h}$ has UV degree of divergence~ $2$, however its UV behaviour is actually milder. In fact, by power-counting the divergent part of $\G^{(1)}_{R\h}$  can be parameterized as
\begin{align}
 \int_x\!\int_y \G^{(1)}_{R_x\h_y} \tR_x \h_y & \underset{\mathrm{UV}\ \mathrm{div}}{=}
\int_x\! \tR_x (c_0^{(1)} + c_1^{(1)} \square ) \h_x.
\end{align}
Nevertheless the coefficient $c_1^{(1)}$ turns out to be zero at one-loop order. This is due to the fact that such a contribution can only be generated by diagrams involving the trilinear derivative interaction $\square X \h^2$. But the differential operator does not act on the external $\h$-leg, so that its action will be that of removing one of the internal propagators, thus leaving a derivative-independent UV divergence.

This is ultimately the reason why there is no BSM contribution to the $p^2$-term in~\1eq{2pt} and hence the wave-function renormalization of the field $\h$ is purely SM.

\subsection{One-loop dimension 6 operator mixing}

Recently the question of which dimension 6 operators might possibly affect the anomalous trilinear Higgs self-coupling has been debated in the literature~\cite{Degrassi:2017ucl}. In addition to the cubic potential term $\Big ( \Phi^\dagger \Phi - \frac{v^2}{2} \Big )^3$ there are two possibilities:
$\partial_\mu (\Phi^\dagger \Phi) \partial^\mu(\Phi^\dagger \Phi)$ and $(\Phi^\dagger D_\mu \Phi)(\Phi^\dagger D^\mu \Phi)$. Both operators would affect the $p^2$-term in the two point function Eq.(\ref{2pt}), which we have already established to be purely SM one.

To begin with observe that, in principle, the operator $\partial_\mu (\Phi^\dagger \Phi) \partial^\mu(\Phi^\dagger \Phi)$ can be generated either from amplitudes involving only external $\h$-legs directly mapped into their counterparts in the target theory, or from amplitudes involving external sources $\tR$ and $\tbarc$ through the mapping~\noeq{repl.rule}. The first type of amplitudes does not contribute: in fact, at $g_6=0$ they are amplitudes of a power-counting renormalizable theory and thus cannot give rise to UV divergences of dimension $6$, as the one possibly associated with $\partial_\mu (\Phi^\dagger \Phi) \partial^\mu(\Phi^\dagger \Phi)$. On the other hand, if one looks at $g_6$-dependent contributions (giving rise to non-power-counting renormalizable Green's functions), one immediately see that no such terms can appear at the one-loop level, as a consequence of the fact that the single BSM interaction vertex is the trilinear $g_6 \Lambda X_2^3$ and the latter only contributes to 1-PI amplitudes with one external $X_2$-leg, which we do not need to consider.

As far as the second type of amplitudes are concerned, since the mapping~\noeq{repl.rule} does not contain derivatives, from the set of 1-PI Green's functions involving external sources we need to consider only those possessing a UV degree $2$, namely $\G^{(1)}_{R\h}$. However, the UV divergence of the latter is momentum-independent, since the coefficient $ c_1^{(1)}$ is zero.
Hence
\begin{align}
 \int_x\!\int_y \G^{(1)}_{R_x\h_y} \tR_x \h_y & \underset{\mathrm{UV}\ \mathrm{div}}{=}
\int_x\! \tR_x (c_0^{(1)} + c_1^{(1)} \square ) \h_x\nonumber \\
& \rightarrow  3 g_6\frac{\Lambda}v\int_x\!\left[ 
  c_0^{(1)} \left(\Phi^\dagger \Phi - \frac{v^2}{2} \right)^2-
c_1^{(1)} \partial_\mu  \left(\Phi^\dagger \Phi\right) \partial^\mu \left(\Phi^\dagger \Phi\right) 
\right] \nonumber\\
& =  3 g_6\frac{\Lambda}v\int_x\!
  c_0^{(1)} \left(\Phi^\dagger \Phi - \frac{v^2}{2} \right)^2,
\end{align}
and indeed the operator $\partial_\mu (\Phi^\dagger \Phi) \partial^\mu(\Phi^\dagger \Phi)$ does not arise. Finally, since the BSM contribution to the $p^2$-term in the two point $\h$ function must vanish, we conclude that also $(\Phi^\dagger D_\mu \Phi)( \Phi^\dagger D_\mu \Phi)$ is not generated.

The vanishing of the coefficient $c_1^{(1)}$ is an intriguing feature of the $X$-theory, whose amplitudes, as already noticed, possess, in general, a milder UV behaviour than one would have expected on the basis of power-counting arguments. Notice in particular that, if $c_1^{(1)}$ were different from zero, one would get a BSM contribution to the $\h$ wavefunction renormalization via the $6g_6\Lambda\G^{(1)}_{R\h}$ term appearing in the first line of~\1eq{2pt}. This would in turn require a {\em polynomial} field redefinition in the target theory; indeed the momentum-squared dependence generated in the two-point $\h$-function by $c_1^{(1)}$ would contribute, under the mapping~\noeq{repl.rule} to Green's functions in the target theory with more than two external $\sigma$-legs. Such contributions are indeed what one would generally expect in a gauge effective field theory, where, as previously remarked, field redefinitions are generic (not necessarily linear) canonical transformations, compatible with the Batalin-Vilkovisky bracket of the model~\cite{Gomis:1995jp}. Thus, it comes as a (pleasant) surprise that they are not present at the one-loop level, though it remains to be seen what happens at higher orders.

\section{\label{sec:arb}Potentials with an arbitrary power ~$X_2^N$}

Besides being illustrative of the advantages of the proposed method, the construction presented in the cubic case lends itself to a generalization for potentials displaying arbitrary powers in the field $X_2$ but no derivatives, namely
\begin{align}
    V(X_2) = \sum_{j=3}^N g_{2j} \Lambda^{4-j} X_2^j,
    \label{pot.1}
\end{align}
where the couplings $g_{2j}$ are dimensionless. Obviously the $X_1$-equation is left unaltered; on the other hand, in order to implement the $X_2$-equation we need additional external sources, namely we set
\begin{align}
	\Gamma_{\tiny{V}}=\int \sum_{j=2}^{N-1} R_{j} X_2^{j},
\end{align}
with $R_2\equiv R$. Then the $X_2$-equation generalizes to
\begin{align}
\G_{X_2} & = \frac{1}{v} \left(\square + m^2\right) \G_{\barc} 
-  \left(\square + m^2\right)X_1   -  \left(\square + M^2\right)X_2 \nonumber \\
& 
+\sum_{j=3}^N\left[
jg_{2j}\Lambda^{4-j}\Gamma_{R_{j-1}}+(j-1)R_{j-1}\Gamma_{R_{j-2}}
\right]
- v \barc, 
\label{X2eq.gen}
\end{align}
where we have defined $\Gamma_{R_1}\equiv X_2$.


Now, the one-loop decomposition of the 1-PI amplitudes introduced in the cubic case stays essentially unchanged also in this more general case. As before, in fact, amplitudes with $\X$-external legs are completely fixed by the $\X$-equations of motion, and again can be traded off for amplitudes involving  insertions of the external sources $\tbarc$ and/or $\tR_j$ with $j\geq2$ (we set ${\mathcal R}\equiv{\mathcal R}_2$, and will identify the sources $\tR_j$ with $j\geq3$ shortly). In addition, the BSM interaction vertices in the $X_2$ potential are at least trilinear, and therefore at the one-loop level the presence of $V(X_2)$ can affect only amplitudes with $X_2$ external legs. The same is evidently true also for any of the $R_j$ sources with $j\geq 3$ whose insertion will generate again diagrams with (at least) an external $X_2$ leg. On the other hand, any $X_2$ dependence can be fully restored by solving the $X_2$-equation of motion and hence amplitudes involving external $X_2$ fields can be discarded. In particular, for UV divergent amplitudes it is sufficient to apply the substitution rule $R_2 \rightarrow \tR_2$, since there are no one-loop UV divergent amplitudes involving $R_j$, $j\geq 3$ at zero $X_2$-fields;  moreover even in this general case {\it amplitudes with $\h$-legs insertions are precisely the same as those of the renormalizable model when the potential $V(X_2)$ is switched off.}

This line of reasoning then implies that the only change with respect to our previous analysis resides in the replacement rule for $R_2$. One can give a general formula for such a replacement. The source of highest index~$N$ is eliminated via the substitution
\begin{align}
	\tR_{N} = R_{N} + (N+1)  \Lambda^{4-(N+1)}g_{2(N+1)}  X_2. 
\end{align}
Then  one proceeds iteratively with the replacement rule being given by
\begin{align}
	\tR_j & = R_j -\sum_{k=1}^{N-j} (-1)^k \frac{(j+1) (j+2) \dots (j+k)}{k!}  
	\nonumber \\ 
	& \times\left[ \Lambda^{4-(j+k)} g_{2(j+k)} +
	(1 - \delta_{j+k, N}) \tR_{j+k} \right] X_2^k;\quad j=2, \dots, N-1.
	\label{repl.gen}
\end{align}
For instance, if operators up to $X_2^4$ are introduced, the $X_2$-equation is  solved by the replacements
\begin{align}
\tR_3 & = R_3+ 4 g_8 X_2 , \quad
\tR_2 = R_2 + 3 (\Lambda g_6 + \tR_3)  X_2 - 6 g_8 X_2^2 ,
\end{align}
with $\tbarc$ as in~\1eq{ext.srcs}. Indeed, if $\Gamma=\Gamma[\tR_3,\tR_2]$, we find
\begin{align}
	\Gamma_{X_2}&\supset \left(\Gamma_{R_3}\frac{\delta R_3}{\delta{\mathcal R}_3}+\Gamma_{R_2}\frac{\delta R_2}{\delta{\mathcal R}_3}\right)\frac{\delta{\mathcal R}_3}{\delta X_2}+\left(\Gamma_{R_3}\frac{\delta R_3}{\delta{\mathcal R}_2}+\Gamma_{R_2}\frac{\delta R_2}{\delta{\mathcal R}_2}\right)\frac{\delta{\mathcal R}_2}{\delta X_2}\nonumber \\
	&=4g_8\Gamma_{R_3}+3\Lambda g_6\Gamma_{R_2}+3R_3\Gamma_{R_2},
\end{align}
which coincides with the part of right-hand side of~\1eq{X2eq.gen} proportional to the derivatives of $\Gamma$ with respect to the external sources.

Notice the appearance of non-linear terms in the $X_2$ field implying that amplitudes with a fixed number of external sources $R_j$ contributes to amplitudes in the target theory with a higher number of $\h$ external legs. For example $\Gamma_{R_2R_2}$ contributes to amplitudes up to four $\h$ legs when only a cubic $X_2$ potential is considered, but to amplitudes up to eight $\h$ legs when a quartic $X_2$ potential is added.  

In fact, it is possible to write down the replacement for $\tR_2$ in closed form; it reads 
\begin{align}
    \tR_2 & = R_2 + \sum_{k=1}^{N-3} c_k X_2^k R_{k+2} 
    + \sum_{k=1}^{N-2} c_{k} \frac{g_{2(k+2)}}{\Lambda^{k-2}} X_2^k, 
    \label{R.map}
\end{align}
where the coefficients $c_k$ are given by
\begin{align}
    c_k = \left( 1 + \left[ \frac{k}{2} \right] \right) \left[ 1 + 
    2 \left( \left[ \frac{k}{2} \right] + (k ~ {\rm mod} ~2) \right) \right],
\end{align}
with $[ x ]$ the greatest integer less or equal than $x$ and $x ~ {\rm mod} ~2$ the remainder on division of $x$ by 2. Surprisingly enough, these coefficients $c_k$ do not depend on $N$ and thus increasing the degree  of the polynomial does not affect lower order terms. Therefore, we can obtain the formula for an arbitrary potential by letting $N \rightarrow \infty$ into~\1eq{R.map}. 

As far as the replacement rule for $\tbarc$ is concerned, it is the same as in \1eq{ext.srcs}, since the $X_1$-equation is left unchanged, and the classical solution~\noeq{sol.X1eom} still holds. On the other hand, the classical $X_2$-equation of motion is modified and reduces to
\begin{align}
    (\square + m^2) (X_1+X_2) = - (M^2-m^2)X_2 + V'(X_2) .
    \label{X1.eom.gen}
\end{align}
where the prime indicates differentiation with respect to $X_2$, we have set to zero all external sources and we made use of ~\1eq{sol.X1eom}. 
This gives the final formula for the mapping
\begin{align}
    \tbarc &= \barc -\frac{M^2 - m^2}{v} X_2 + \frac{1}{v} V'(X_2)\nonumber \\
    &=\barc +\frac1v\sum_{j=2}^Njg_{2j}\Lambda^{4-j}X_2^{j-1},
      \label{repl.rule.barc.gen}  
\end{align}
where we have set $2\Lambda^2g_{4}=m^2-M^2$. Notice that again we can let $N\to\infty$.

Thus let us recap. Consider a generic potential of the type~\noeq{pot.1}, possibly with $N\to\infty$. At one-loop, the divergent amplitudes in the $\X$-theory are the same eleven ones identified in the cubic case $N=3$, and explicitly evaluated in Appendix~\ref{app:cts}. Then, by applying to these latter amplitudes the mappings~\noeq{R.map} (at zero sources $R_j=0$,  $j\ge3$) and~\noeq{repl.rule.barc.gen} and using the classical $X_1$-equation of motion, \ie by substituting $X_2$ with~\1eq{sol.X1eom}, one can generate, {\it in a purely algebraic way}, all the divergent amplitudes in the $\h$-sector of the target theory.

\section{\label{sec:renorm}One-loop off-shell algebraic renormalization}

The results obtained in the previous section allows us to carry out the off-shell one-loop renormalization of the Higgs sector for an arbitrary potential $V$ of the type~\noeq{pot.1}, within the framework of algebraic renormalization. 

Let us remind the reader that algebraic renormalization~\cite{Piguet:1995er,Becchi:1974md,Becchi:1975nq} is a regularization scheme-independent technique that allows one to study the renormalization of gauge theories in a mathematically rigorous way by exploiting the  locality properties of quantum field theory (encoded in the so-called Quantum Action Principle~\cite{Zimmermann:1969jj,Zimmermann:1972te,Breitenlohner:1977hr}) together with powerful cohomological tools rooted into the nilpotency of the BRST differential (for a review see {\it e.g.},~\cite{Barnich:2000zw}). It allows to classify the action-like counterterms as well as the anomalies (or lack thereof) of the model, and  has been used in a variety of phenomenological applications, {\it e.g.}, in order to establish the gauge-independence to all orders in the loop expansion of the pole mass of physical fields in the SM~\cite{Gambino:1999ai}, and to study the gauge invariance for fermion mixing renormalization~\cite{Gambino:1998ec}. Use of non-invariant regularization schemes and the derivation of the associated finite counterterms, restoring the relevant symmetries of the theory, have been discussed in~\cite{Grassi:1999tp} for the QCD corrections to the Higgs decay into two photons and to two-loop electroweak corrections to $B\rightarrow X_s \gamma$, and in~\cite{Grassi:2000kp} for the non-invariant two-loop counterterms for the SM three-gauge-boson vertices. Other results include: the study of the background field method applied to the process $b \rightarrow s\gamma$~\cite{Grassi:2001zz}, the constraints imposed on the IR behaviour of Yang-Mills Green's functions in the Landau gauge~\cite{Grassi:2004yq}, the derivation of a general scheme-independent technique for describing the action-like sector of a gauge theory fulfilling the defining Slavnov-Taylor identities of the model~\cite{Ferrari:1999nj,Quadri:2003ui,Quadri:2003pq,Quadri:2005pv}, and, the one-loop renormalization of a general chiral gauge in the presence of a non-anticommuting $\gamma_5$~\cite{Martin:1999cc}. The renormalization of the SM to all orders has been presented in~\cite{Kraus:1997bi,Grassi:1999nb}, the Minimal Supersymmetric Standard Model has been studied in~\cite{Hollik:2002mv}, and, finally, models based on non-linearly realized symmetries have been discussed in~\cite{Ferrari:2005va,Ferrari:2005fc,Bettinelli:2007kc,Bettinelli:2007tq,Bettinelli:2007cy,Bettinelli:2008ey,Bettinelli:2008qn,Bettinelli:2009wu}.

\subsection{Summary of results}

Before dwelling into the details, let us, for reference purposes, collect below the relevant equations of the model derived so far. 
\begin{itemize}
	\item {\bf Action.} The classical action is given by 
\begin{align}
	\Gamma=\Gamma_\s{\mathrm{SM}}+\Gamma_{\tiny{\mbox{SSB}}}+\Gamma_{\tiny{\mbox{SRC}}}, 
\end{align}
where $\Gamma_\s{\mathrm{SM}}$ is the usual SM action including the Yang-Mills, fermion, Yukawa, (linear) gauge fixing and ghost terms; the spontaneous symmetry-breaking part $\Gamma_{\tiny{\mbox{SSB}}}$ reads
\begin{align}
    \Gamma_{\tiny{\mbox{SSB}}}&=\int
    \left[D_\mu\Phi^\dagger D^\mu\Phi-\frac{M^2-m^2}2 X_2^2
    -\frac{m^2}{2v^2} \Big ( \Phi^\dagger\Phi-\frac{v^2}2 \Big )^2-\overline{c}\left(\square+m^2\right) c
    \right.\nonumber \\
    &
    +\left.\frac1v\left(X_1+X_2\right)(\square+m^2)\left(\Phi^\dagger\Phi
    -\frac{v^2}2-vX_2
    \right)+ \sum_{j=3}^\infty g_{2j} \Lambda^{4-j} X_2^j\right],
    \label{action.recap}
\end{align}
and contains the arbitrary analytic potential in the last term.
Finally the external source sector, including the
source $\barc$ needed to formulate the $X_1$-equation of motion and  
the sources $R_j$ necessary to derive the $X_2$-equation of motion when
the full towers of higher order operators $X_2^j$ is switched on is 
\begin{align}
\Gamma_{\tiny{\mbox{SRC}}} = \int\overline{c}^*\left(\Phi^\dagger\Phi-\frac{v^2}2-vX_2\right)+\int \sum_{j=2}^{\infty} R_{j} X_2^{j}.
\end{align}

\item {\bf $X$-equations.} The $X_1$-equation of motion is the same as in the cubic case, namely
\begin{align}
    \G_{X_1}& =\frac{1}{v} \left(\square + m^2\right) 
    \G_{\barc}. 
    \label{recap.X1eom}
 \end{align}
The $X_2$-equation of motion becomes instead
\begin{align}
  \G_{X_2} & = \frac{1}{v} \left(\square + m^2\right) \G_{\barc} 
  -  \left(\square + m^2\right)X_1   -  \left(\square + M^2\right)X_2 \nonumber \\
  & 
  +\sum_{j=3}^\infty\left[
  jg_{2j}\Lambda^{4-j}\Gamma_{R_{j-1}}+(j-1)R_{j-1}\Gamma_{R_{j-2}}
  \right]
  - v \barc.
\label{X2eom.recap}
\end{align}

\item {\bf Mappings.} \1eq{recap.X1eom} is solved to all order in the loop expansion by the replacement
\begin{align}
    \tbarc &= \barc + \frac{1}{v}\left(\square + m^2\right)\left(X_1+X_2\right).
\end{align}
In the one-loop approximation the tree-level equation of motion for $X_2$
can be used in order to substitute the rhs of the above equation in order to eliminate 
the Klein-Gordon operator, as follows:
\begin{align}
    \tbarc &=\barc +\frac1v\sum_{j=2}^Njg_{2j}\Lambda^{4-j}X_2^{j-1}.
\end{align}
On the other end the all-order solution to Eq.(\ref{X2eom.recap}) can be iteratively 
reconstructed via Eq.(\ref{repl.gen}).
In particular the explicit solution for the $R_2$ source (the only relevant one if 
one is interested in the UV divergent amplitudes of the target theory) is given by
\begin{align}
       \tR_2 & = R_2 + \sum_{k=1}^{N-3} c_k X_2^k R_{k+2} 
    + \sum_{k=1}^{N-2} c_{k} \frac{g_{2(k+2)}}{\Lambda^{k-2}} X_2^k, 
    \label{R.map.recap}
\end{align}
with
\begin{align}
    c_k = \left( 1 + \left[ \frac{k}{2} \right] \right) \left[ 1 + 
    2 \left( \left[ \frac{k}{2} \right] + (k ~ {\rm mod} ~2) \right) \right],
\end{align}
$[x]$ being the greatest integer less or equal than $x$ and $x ~\mbox{mod } 2$
the remainder on division of $x$ by $2$.
\end{itemize}

\subsection{$\X$-theory}

Let us consider first how one-loop renormalization works in the $\X$-theory. To begin with observe that the renormalization program needs to be carried out  only in the $\X$-independent sector, since amplitudes involving $X_1$ and/or $X_2$ insertions are generated, at this level, in a purely algebraic way   through the replacements~\noeq{R.map} and~\noeq{repl.rule.barc.gen} at zero external sources $R_j$ (recall that one-loop 1-PI amplitudes involving at least one such source with $j\geq 3$ will have at least an external $X_2$ leg). Finally, without loss of generality, we can consider the cubic case only, as at one-loop the $\X$-independent sector is not sensitive to the presence of terms of dimension $>4$ in the potential $V$; for this case, all the divergent 1-PI amplitudes have been calculated in Appendix~\ref{app:cts}.  

 Following standard Algebraic Renormalization techniques, one can perform the expansion of the UV divergences of the theory at one loop order on a basis of BRST invariants, which in the $\X$-independent sector can be constructed from $R$, $\barc$ and $\Phi$ only. There are eleven such invariants:
\begin{align}
	&\int\!\c{1}\barc;& &\int\!\frac12\c{2}\bar c^{*2};&  &\int\! \c{3}R,& \nonumber \\ 
	&\int\!\frac12\c{4}R^2; & &\int\! \c{5}\barc R;& 
	&\int\!\c{6}R\left(\Phi\Phi^\dagger-\frac{v^2}2\right),&\nonumber \\
	&\int\!\c{7}R\left(\Phi\Phi^\dagger-\frac{v^2}2\right)^2;&  
	&\int\!\frac12\c{8}R^2\left(\Phi\Phi^\dagger-\frac{v^2}2\right);&
	&\int\!\frac12\c{9}R^2\left(\Phi\Phi^\dagger-\frac{v^2}2\right)^2,\nonumber\\
	&\int\!\c{10} \barc\left(\Phi\Phi^\dagger-\frac{v^2}2\right);&
	&\int\!\c{11}R\bar c^*\left(\Phi\Phi^\dagger-\frac{v^2}2\right).
	\label{invariants}
\end{align}

Start then from the $\h$-independent sector. There are five independent divergent amplitudes in this sector, see \1eq{uv.div.barcR}, which can be  trivially reabsorbed through the first five invariants above when
\begin{align}
 \c{1} &= \frac{1}{16 \pi^2}\frac{1}{\epsilon}
\left( 2 \MW^2 + \MZ^2 + M^2 \right);& 
\c{2} &=  -\frac{1}{4 \pi^2}\frac{1}{\epsilon}; &
\c{3} &= \frac{M^2}{8 \pi^2}\frac{1}{\epsilon},  \nonumber \\
 \c{4} &= -\frac{1}{4 \pi^2} \frac{1}{\epsilon};&
\c{5} &= -\frac{1}{8 \pi^2} \frac{1}{\epsilon}.& 
\label{inv.1}
\end{align}

Next consider the $R\sigma$-sector, with its four divergent amplitudes, given by the left terms in \1eq{uv.div.Rsigma}. The sixth and seventh invariants in~\1eq{invariants} lead to the Higgs monomials 
\begin{align}
    \int\left[\c{6}R\h+\left(\frac1{2v}\c{6}+\c{7}\right)R\h^2+\frac1v\c{7}R\h^3+\frac1{4v^2}\c{7}R\h^4\right].
\end{align}
Taking into account the appropriate combinatorial factors it is easily seen that the counterterms for the four divergent amplitudes are given by
\begin{align}
    \c{6}=\frac{m^2 + 4 M^2}{8\pi^2v^2}\frac1{\epsilon}; \qquad
    \c{7}=\frac{m^2 + 2 M^2}{4\pi^2v^4}\frac1{\epsilon}.
\end{align}

A similar pattern persists in the $R^2\sigma$ sector, whose divergences are presented in the right terms of \1eq{uv.div.Rsigma}. In this case one has to look at the eighth and ninth invariant in~\noeq{invariants} which give rise to the Higgs monomials 
\begin{align}
    \int\left[\frac12\c{8}R^2\h+\frac12\left(\frac1{2v}\c{8}+\frac12\c{9}\right)R^2\h^2+\frac1{2v}\c{9}R^2\h^3+\frac1{8v^2}\c{9}R^2\h^4\right].
\end{align}
Again after taking into account the appropriate combinatorial factors, the choice 
\begin{align}
    \c{8}&=-\frac1{\pi^2v^2}\frac{1}{\epsilon};\qquad
    \c{9}=-\frac1{\pi^2v^4}\frac{1}{\epsilon},
\end{align}
reabsorbs all divergent terms.

Next consider the $\bar c^*\sigma$-sector, whose two divergent amplitudes are given in~\1eq{uv.div.barch}. The tenth invariant in~\noeq{invariants} is the one needed in this case; however, there is also a tree-level coupling between $\bar c^*$ and $\h$ to take into account, so that field redefinitions play a role in this sector.

Yet, recall that~\1eq{2pt} tells us that the wave function renormalization of the $\h$ field is the same as the SM one; this implies in turn that the renormalization of the Higgs field will happen only through the SM-like $\h$-field redefinition $\h'=Z_\h\h$, and that the possibility of having a non-linear field redefinition of $\h$ does not materialize.  

Thus the relevant contributions in this sector are (as usual we neglect terms depending on the Goldstone bosons)
\begin{align}
    \int \left[\c{10} \barc\left(\Phi\Phi^\dagger-\frac{v^2}2\right) +
    \bar c^*\left(\frac12Z^2_\h\h^2+(v+\delta v) Z_\h \h\right)
    \right].
\end{align}
Setting $Z_\h=1+\delta\h$, we get in the one-loop approximation
\begin{align}
    \int \left[
    \left(\delta v + v\c{10}+v\delta\h\right)\bar c^*\h+
    \frac12\left(\c{10}+2\delta\h\right)\bar c^*\h^2
    \right].
\end{align}

This yields the result
\begin{align}
	\c{10}=- \Gamma^{(1)}_{\bar c^*\h\h}-2\delta\h,
\end{align}
together with the following consistency condition prescribing the renormalization of the vev:
\begin{align}
	\delta v-v\delta\h=-\Gamma^{(1)}_{\bar c^*\h} + v\Gamma^{(1)}_{\bar c^*\h\h}.
	\label{consistency}
\end{align}
Notice that the above equation predicts the renormalization of the vev in terms of
two contributions: the first one in the lhs is related to the wave-function renormalization of the scalar field, the second one is an extra term governed by the external source $\barc$. This provides an alternative representation of the gauge-invariant vev renormalization decomposition given in~\cite{Sperling:2013eva}.

Now, the renormalization of the $\sigma$ field in the conventions of~\cite{Denner:1991kt} is
\begin{align}
    \delta\h&= 2 \partial_{q^2}\Gamma^{(1)}_{\h\h}(q^2),
\end{align}
with an explicit evaluation yielding (we set $\alpha=e^2/4\pi$)
\begin{align}
	\delta\h&= \frac{\alpha}{8\pi \sw^2\cw^2}\frac1\epsilon(1+2\cw^2)-\frac{\alpha}{8\pi\sw^2\MW^2}\frac1\epsilon\left(\sum_{\ell}m^2_\ell+3\sum_{q}m^2_q\right).
\end{align}
Next, using~\1eq{consistency}, we obtain
\begin{align}
 \delta v&= \frac{\alpha\MW}{2\pi e\sw\cw^2}\frac1\epsilon(1+2\cw^2)
    -\frac{\alpha}{4\pi e\sw\MW}\frac1\epsilon\left(\sum_{\ell}m^2_\ell+3\sum_{q}m^2_q\right).
    \label{pred.deltav}
\end{align}
Notice that this result coincides to the renormalization of the tree-level expression for \mbox{$v^2=\frac{4 M^2_W s^2_W}{e^2}$}, namely
\begin{align}
    \delta v = v\left(\frac{\delta\sw}{\sw}+\frac{\delta\MW^2}{2\MW^2}-\delta Z_e\right)
\end{align}
when one takes into account the relevant SM renormalization constants given in~\cite{Denner:1991kt}:
\begin{align}
    \delta\MW^2&=-\Gamma^{(1)}_\s{WW}(q^2)\Big|_{q^2=\MW^2},\nonumber \\
    \delta Z_e &=-\frac12\partial_{q^2}\Gamma^{(1)}_\s{\gamma\gamma}(q^2)\Big|_{q^2=0}+\frac{\sw}{\cw\MZ^2}\Gamma^{(1)}_\s{\gamma Z}(q^2)\Big|_{q^2=0},\nonumber \\
   \delta\sw&=\frac12\frac{\cw^{2}}{\sw}\left(\frac1{\MW^2}\Gamma^{(1)}_\s{WW}(q^2)\Big|_{q^2=\MW^2}-\frac1{\MZ^2}\Gamma^{(1)}_\s{ZZ}(q^2)\Big|_{q^2=\MZ^2}\right),
\end{align}
where for gauge bosons  only the transverse part of the self-energy enters in the relations above.


Putting all together we then have the result  
\begin{align}
	\c{10}=-\frac{\alpha}{8\pi \sw^2\cw^2}\frac1\epsilon\left[3(1+2\cw^2)- \cw^2 \frac{2 m^2 + M^2}{\MW^2}\right]+\frac{\alpha}{4\pi\sw^2\MW^2}\frac1\epsilon\left(\sum_{\ell}m^2_\ell+3\sum_{q}m^2_q\right). 
\end{align}

The final sector to be considered is the mixed one $R\bar c^*\sigma$, in which the two divergent amplitudes~\noeq{uv.div.Rbarch} ought to be reabsorbed by the last of the invariants~\noeq{invariants} 
\begin{align}
    \int\left[
    \c{11}v R\bar c^*\h +
    \frac{1}{2}\c{11} R\bar c^*\h^2\right].
\end{align}
Indeed this happens when choosing
\begin{align}
   \c{11} = -\frac{1}{4 \pi^2 v^2} \frac{1}{\epsilon}. 
\end{align}

Thus, the eleven BRST-invariant counterterms $\c{i}$ together with the usual linear SM $\h$-field redefinition $\h'=Z_\h\h$ and the vev renormalization $v\to v+\delta v$ allows to reabsorb all the one-loop divergences of the $\X$-theory in the cubic potential case $N=3$, a result that, as explained above, generalizes to any $N$.

\subsection{Target theory}

Renormalization of the target theory can be successfully achieved in a rather straightforward way: one imposes in the replacement rules~\noeq{R.map} and~\noeq{repl.rule.barc.gen} (at zero external sources) the classical $X_1$-equation of motion, \ie by substitutes $X_2 \rightarrow \frac{1}{v}\left ( \Phi^\dagger \Phi - \frac{v^2}{2} \right )$. Thus, in the target theory the BSM potential is introduced directly at tree-level as 
\begin{align}
	V = \sum_{j=3}^\infty g_{2j}  \frac{\Lambda^{4-j}}{v^j} \left ( \Phi^\dagger \Phi - \frac{v^2}{2} \right )^j,
	\label{bsm.pot.target}
\end{align}
and renormalization requires to find out the appropriate field, vev and coupling constant renormalizations absorbing all one-loop divergences. Notice that all the coupling constants $g_{2j}$ will be renormalized, \ie one needs to determine an infinite number of counterterms already at the one-loop order. 

The results achieved in the $\X$-theory case tells us however that only a handful of these counterterms are independent; more specifically, in order to control the infinite number of divergences arising in the target theory one needs to consider only the eleven coefficients  $\c{i}$ of the BRST invariants determined previously, supplemented with the Higgs wave function and vev renormalizations.




Let's first consider the $\h$ and $v$ renormalization applied on the tree-level $\barc$-dependent terms in
the $\X$-theory; they yield under the mapping~\noeq{repl.rule.barc.gen} for $\barc$
\begin{align}
& \int \! \barc \left [ ( \delta v + v \delta \sigma) \sigma + \delta \sigma \sigma^2\right] \rightarrow \int\!\frac1{v^2}\left.\sum_{j=2}^\infty jg_{2j}\Lambda^{4-j}X_2^{j-1}\right |_{X_2 = \frac{1}{v} \left(\Phi^\dagger \Phi - \frac{v^2}{2} \right)} \left [ ( \delta v + v \delta \sigma) \sigma + \delta \sigma \sigma^2\right ].
\label{wfv.ren.frombarc}
\end{align}
Notice that the $m^2$-contribution contained in the $g_4$ term exactly cancels against the $\sigma$ and $v$ renormalization of the Higgs potential in the $\X$-theory~\1eq{action}.

Now, in the target theory the wave-function and vev renormalization of the Higgs and the BSM potential~\noeq{bsm.pot.target} give two contributions: the first is induced by the wave-function and vev renormalization of the invariant $\Phi^\dagger \Phi - \frac{v^2}{2}$ and matches exactly~\1eq{wfv.ren.frombarc} above at $m=0$; the second is generated by the vev renormalization of the $v$-dependence of the coefficients of the monomials in $\Phi^\dagger \Phi - \frac{v^2}{2}$, and gives 
\begin{align}
     \int\! \left [ 2 \frac{M^2}{v^3}  \left ( \Phi^\dagger \Phi - \frac{v^2}{2} \right )^2 - \sum_{j=3}^\infty  j g_{2j}  \frac{\Lambda^{4-j}}{v^{j+1}} \left ( \Phi^\dagger \Phi - \frac{v^2}{2} \right )^j \right ] \delta v.
     \label{g2j.vev}
\end{align}
The crucial point is that the above contribution is gauge-invariant and thus can be reabsorbed by the redefinition of the coefficients $M$ and $g_{2j}$. We can thus conclude that the non-gauge-invariant operators arising from the vev and field redefinitions are automatically taken into account through the mapping of the external source $\tbarc$.

Finally, it is possible to give a closed analytical form for the renormalization of the BSM coupling constants $g_{2j}$ by considering what happens to the UV divergent BRST-invariants of the $\X$-theory under the mappings ~\noeq{R.map} and~\noeq{repl.rule.barc.gen} and projecting the result onto the monomial $\Lambda^{4-j}X_2^j=\frac{\Lambda^{4-j}}{v^j} \left ( \Phi^\dagger \Phi - \frac{v^2}{2} \right )^j$. 

For example, considering the first two invariants in~\1eq{invariants}, one finds
\begin{align}
	\c{1}\barc\to\c{1}\frac1v\sum_{k=2}^\infty kg_{2k}\Lambda^{4-k}X_2^{k-1}\underset{k=j+1}{\to}\c{1}\frac{j+1}{\Lambda v}g_{2(j+1)}\Lambda^{4-j}X_2^j\quad\Rightarrow\quad \delta g_{2j} \supset \c{1}\frac{j+1}{\Lambda v}g_{2(j+1)},
\end{align} 
and
\begin{align}
	\frac12\c{2}\bar c^{*2}&\to\c{2}\frac1{2v^2}\sum_{k=2}^\infty\sum_{\ell=2}^\infty k\ell g_{2k}g_{2\ell}\Lambda^{8-k-\ell}X_2^{k+\ell-2}\nonumber \\
	&\hspace{-.4cm}\underset{k+\ell-2=j}{\to}\c{2}\frac{\Lambda^2}{2v^2}\sum_{k =2}^{j+2-k\ge2} k ( j + 2 - k) g_{2k} g_{2(j+2-k)}\Lambda^{4-j}X_2^j\nonumber \\
	&\Rightarrow  \delta g_{2j} \supset \c{2}\frac{\Lambda^2}{2v^2}\sum_{k =2}^{j} k ( j + 2 - k) g_{2k} g_{2(j+2-k)}.
\end{align}
Proceeding in this way for all the remaining invariants one then finds the final expression
\begin{align}
   \delta g_{2j} & = \c{1} \frac{j+1}{\Lambda v} g_{2(j+1)} + 
   \c{2} \frac{\Lambda^2}{2v^2} \sum_{k =2}^j k ( j + 2 - k) g_{2k} g_{2(j+2-k)}+
   \c{3} \frac{c_j}{\Lambda^{2}}  g_{2(j+2)} \nonumber \\
   & + \c{4}\frac{1}{2} \sum_{k = 1}^{j-1} c_k c_{j-k} g_{2(k+2)} g_{2(j-k+2)}  +
   \c{5} \frac{\Lambda}{v} \sum_{k=2}^{j} k c_{j+1-k} g_{2k} g_{2 (j+3-k)} \nonumber \\
   & + \c{6}\frac{v}{\Lambda} c_{j-1} g_{2(j+1)} + \c{7} v^2 c_{j-2} g_{2j} \nonumber \\
   & + \c{8} \frac{\Lambda v}{2} \sum_{k=1}^{j-2} c_k c_{j-k-1}
   g_{2(k+2)} g_{2(j-k+1)} +
   \c{9} \frac{\Lambda^2 v^2}2 \sum_{k=1}^{j-3} c_k c_{j-k-2} g_{2(k+2)} g_{2(j-k)} \nonumber \\
   & + \c{10} j g_{2j} + \c{11} \Lambda^2 
    \sum_{k=1}^{j-2} (j-k) c_k g_{2(k+2)} g_{2(j-k)}  - \frac{j}{v} \delta v,
     \label{g2j.Rsigma}
\end{align}
where the last term comes from subtracting the contribution from the vev renormalization in~\1eq{g2j.vev}, and the combinatorial factors $c_k$ are given in~\1eq{R.map}. 

Notice that the renormalized $g_{2j}$ receives contributions from higher order coupling constants $g_{2\ell}$, $\ell\geq j$. This is a well-known fact in the literature (see for example~\cite{DiLuzio:2017tfn}), and it constitutes in fact the reason why in the target theory we have considered the complete potential, containing the full tower of $\Phi^\dagger\Phi$ operators. On the other hand~\1eq{g2j.Rsigma} gives one full control on the renormalization of the full tower of operators entering in the analytical potential $V$, so that the one-loop stability problem can be addressed via a complete evaluation of the $\beta$-functions for the coupling constants $g_{2j}$. This is however beyond the scope of the present paper and will be discussed elsewhere.

We conclude observing that the renormalized couplings in the target theory must not depend on the mass parameter $m^2$, since the latter is an unphysical quantity. Indeed one can explcitly verify that the dependence on $m^2$, arising from the coefficients $\c{i}$, does indeed cancel out for every $j$, a result that provides a strong consistency check of the above formula.

\section{\label{sec:concl}Conclusions}

We have extended the algebraic renormalization program to the case of the one-loop renormalization of HEFTs possessing an arbitrary analytic derivative-independent BSM potential depending only on the gauge singlet~$\Phi^\dagger \Phi-\frac{v^2}2$.
Has as been emphasized in the paper, this is a highly non-trivial task, since in a conventional approach an infinite number of divergent counterterms arise already at one-loop, and, in addition, the issue of controlling the field redefinitions is cumbersome, to say the least, since one cannot anymore use power-counting arguments in order to constrain the field redefinitions allowed by the symmetries of the theory.

Contrary to naive expectations, we have shown that in a particular reformulation of the SSB mechanism by means of suitable $X$-auxiliary fields controlled by an extended BRST symmetry,  some further functional identities hold, strongly constraining the UV divergences of the theory. Indeed, in the scalar sector of the $X$-theory there are only eleven UV divergent independent invariants. The latter involve the external sources $\barc$ and $R$ and the gauge singlet $\Phi^\dagger \Phi-\frac{v^2}2$.
On the other hand, Green's functions involving the $X_2$-fields are recovered by a purely algebraic technique via the substitution rules for the external sources, solving the $X_{1,2}$-equations of motion. These substitution rules are valid to all orders in the loop expansion. Then, in order to recover the divergences of the scalar sector of the target theory one simply needs to go on-shell with the $X$-fields. At one loop-order the prescription is particularly straightforward and amounts to replace $X_2$ with the gauge singlet in both the $\tbarc$ and $\tR$ substitution rules.

Let us then summarize the main results. One finds that the wave-function renormalization of the $\h$ field is the purely SM one; in addition, the wave-function renormalization contributions in the target theory are automatically taken into account by the $\tbarc$-substitution rule. Then one can disentangle the genuine contribution to the gauge-invariant operators
$\left ( \Phi^\dagger \Phi-\frac{v^2}2 \right )^j$, obtaining in closed form the one-loop renormalization of the BSM coupling constants $g_{2j}$. These results hold for any analytic potential $V$ depending on arbitrary powers of $\Phi^\dagger \Phi-\frac{v^2}2$.

One can envisage several applications. Beisde the aforementioned computation of the $\beta$-functions of the BSM coupling constants which is currently under way, one has the opportunity to study the higher order renormalization of the theory. While one expects more and more new independent divergences to appear, in the spirit of the HEFTs, the higher order constraints arising from the functional identities in the $X$-theory are in fact an interesting subject that awaits to be studied. Finally, the $\X$-theory is potentially applicable to cosmological relaxation theories~\cite{Graham:2015cka}, in which the Higgs mass is stabilized through classical dynamics. In this context, the sources $\tR$ seems to be the right tool through which one can describe a quantized field in the presence of external sources and a (derivative-independent) arbitrary potential. 

We hope to come back to these issues in the near future.

\section*{Acknowledgments}

Useful discussions with Heidi Rzehak and Tilman Plehn are gratefully acknowledged. A.Q. thanks the European Centre for Theoretical Studies in Nuclear Physics and Related Areas (ECT*), where part of this work has been carried out under an agreement with INFN, and the Institute for Theoretical Physics at the Heidelberg University for the warm hospitality.

\appendix

\section{Explicit results for the one-, two- and three-point $\sigma$ amplitudes}\label{app:3sigma}

From~\2eqs{1pt}{2pt} we get for the one- and two-point Higgs functions in the target theory
\begin{align}
    \widetilde{\G}^{(1)}_{\h}&\underset{g_6\ \mathrm{terms}}{=}\frac{3\Lambda}{16\pi^2}g_6A_0(M^2), \\
    \widetilde{\G}^{(1)}_{\h\h}&\underset{g_6\ \mathrm{terms}}{=}\frac{9\Lambda^2}{8\pi^2}g_6^2B_0(p^2,M^2,M^2)+\frac{3\Lambda}{16\pi^2v}g_6\left[6A_0(M^2)\right.\nonumber \\
    &\left.+A_0(M^2_\s{Z})+2A_0(M^2_\s{W})+6M^2B_0(p^2,M^2,M^2)\right],
\end{align}
where $A_0$ and $B_0$ denote the one- and two- point Passarino-Veltman  (PaVe) scalar functions\footnote{We use the notation of~\cite{Denner:1991kt}.}.

For the three-point amplitude $\widetilde{\G}^{(1)}_{\h\h\h}$ one can proceed as described in Section~\ref{map}, obtaining the result
\begin{align}
    \tG^{(1)}_{\sigma\sigma\sigma}  &\underset{g_6\ \mathrm{terms}}{=}
    - \frac{27\Lambda^3}{2 \pi^2}g_6^3  C_0(p_1^2,p_2^2,p_3^2)\nonumber \\
    & + \frac{27 \Lambda^2}{4 \pi^2v}g_6^2 \left[ M^2 C_0(p_1^2,p_2^2,p_3^2) + B_0(p_1^2,M^2,M^2) + \mathrm{cyclic} \right] \nonumber \\
    & -\frac{3M^2\Lambda}{16 \pi^2v^2}g_6 \left[ 18 M^2 C_0(p_1^2,p_2^2,p_3^2)+ 21  B_0(p_1^2,M^2,M^2)\right.
    \nonumber \\
    & \left. +  B_0(p_1^2,\MZ^2,\MZ^2) + 2 B_0(p_1^2,\MW^2,\MW^2) + \mathrm{cyclic} 
    \right] \nonumber \\ 
    &- \frac{9\Lambda}{16 \pi^2 v^2}g_6 \left[ 2 A_0(\MW^2) + A_0(\MZ^2) + 5 A_0(M^2) \right],
\label{3pt}
\end{align}
where $C_0(p_1^2,p_2^3,p_3^2) \equiv C_0(p_1^2,p_2^3,p_3^2;M^2,M^2,M^2)$
is the  three-point PaVe scalar function of equal mass $M^2$, and, where indicated, we cyclically sum over the three momenta $p_1,\ p_2$ and~$p_3$.

Notice finally that, as expected, in all the results no dependence on the unphysical mass parameter $m$ is present.

\section{$g_6$-dependent one-loop counterterms}\label{app:cts}

In this appendix we derive all the counterterms needed to renormalize the Higgs sector of a BSM target theory with a sextic Higgs potential, using the power counting rules introduced in Section~\ref{powerc}.

To begin with, observe that amplitudes involving only external $\sigma$-legs can be neglected since in the $\X$ theory they never contribute to $g_6$-dependent terms. Next, there are five divergent amplitudes with external sources insertions and no $\sigma$-external legs ($\epsilon=4-D$ where $D$ is the space-time dimension):
\begin{align}
& - \G^{(1)}_{\barc} \underset{\mathrm{UV}\ \mathrm{div.}}{=} \frac{1}{16 \pi^2}\frac{1}{\epsilon}
\left( 2 \MW^2 + \MZ^2 + M^2 \right), \nonumber \\
& - \G^{(1)}_{\barc\barc} \underset{\mathrm{UV}\ \mathrm{div.}}{=}  -\frac{1}{4 \pi^2}\frac{1}{\epsilon};
\qquad
 - \G^{(1)}_{R} \underset{\mathrm{UV}\ \mathrm{div.}}{=} \frac{M^2}{8 \pi^2}\frac{1}{\epsilon},
 \nonumber \\
& -\G^{(1)}_{R\barc} \underset{\mathrm{UV}\ \mathrm{div.}}{=}  -\frac{1}{8 \pi^2} \frac{1}{\epsilon};
\qquad -\G^{(1)}_{RR} \underset{\mathrm{UV}\ \mathrm{div.}}{=}  -\frac{1}{4 \pi^2} \frac{1}{\epsilon} .
\label{uv.div.barcR}
\end{align}

Finally, we need to consider amplitudes involving the external sources and $\sigma$-insertions, which are of three types:
\begin{enumerate}
\item Two amplitudes involving $\barc$-insertions and no $R$-legs:
\begin{align}
& - \G^{(1)}_{\barc\h} \underset{\mathrm{UV}\ \mathrm{div.}}{=}
\frac{1}{8 \pi^2 v} \frac{1}{\epsilon} \left(2 m^2+ M^2\right), \nonumber \\
& - \G^{(1)}_{\barc\h\h} \underset{\mathrm{UV}\ \mathrm{div.}}{=} 
-\frac{1}{8 \pi^2 \cw^2 v^2} \frac{1}{\epsilon} 
\left[\left( 1 + 2 \cw^2\right)\MW^2 - \cw^2 \left(2 m^2 + M^2\right) \right]. 
\label{uv.div.barch}
\end{align}

\item Eight amplitudes with $R$-insertions and no $\barc$-legs:
\begin{align}
& - \G^{(1)}_{R\h} \underset{\mathrm{UV}\ \mathrm{div.}}{=} 
\frac{1}{8 \pi^2 v}\frac{1}{\epsilon}  \left(m^2 + 4 M^2\right); &  
& - \G^{(1)}_{R R \h} \underset{\mathrm{UV}\ \mathrm{div.}}{=} 
-\frac{1}{\pi^2 v}\frac{1}{\epsilon} ,\nonumber \\
& - \G^{(1)}_{R\h\h} \underset{\mathrm{UV}\ \mathrm{div.}}{=} 
\frac{1}{8 \pi^2 v^2}\frac{1}{\epsilon}  \left(5m^2 + 12 M^2\right); &
&  - \G^{(1)}_{R R \h \h} \underset{\mathrm{UV}\ \mathrm{div.}}{=} 
-\frac{3}{\pi^2 v^2}\frac{1}{\epsilon}, \nonumber \\
& - \G^{(1)}_{R\h\h\h} \underset{\mathrm{UV}\ \mathrm{div.}}{=} 
\frac{3}{2 \pi^2 v^3}\frac{1}{\epsilon}  \left(m^2 + 2 M^2\right); &
&  - \G^{(1)}_{R R \h \h \h} \underset{\mathrm{UV}\ \mathrm{div.}}{=} 
-\frac{6}{\pi^2 v^3}\frac{1}{\epsilon}, \nonumber \\
& - \G^{(1)}_{R\h^4} \underset{\mathrm{UV}\ \mathrm{div.}}{=} 
\frac{3}{2 \pi^2 v^4}\frac{1}{\epsilon}  \left(m^2 + 2 M^2\right); &
&  - \G^{(1)}_{R R \h^4} \underset{\mathrm{UV}\ \mathrm{div.}}{=} -\frac{6}{\pi^2 v^4}\frac{1}{\epsilon}.
\label{uv.div.Rsigma}
\end{align}

\item Two mixed $R$-$\barc$-amplitudes:
\begin{align}
& - \G^{(1)}_{R\barc\h} \underset{\mathrm{UV}\ \mathrm{div.}}{=} 
-\frac{1}{4 \pi^2 v} \frac{1}{\epsilon}; &
& - \G^{(1)}_{R\barc\h\h} \underset{\mathrm{UV}\ \mathrm{div.}}{=} 
-\frac{1}{4 \pi^2 v^2} \frac{1}{\epsilon}.
\label{uv.div.Rbarch}
\end{align}

\end{enumerate}

Then, from the mapping in \1eq{repl.rule} we see that each $\tR$ (respectively, $\tbarc$) can contribute up to two (respectively, four) $\h$-insertions; thus from the above list we conclude that the one-loop UV-divergent amplitudes in the target theory, involving $\h$-legs only, have at most eight $\h$-insertions. The required BSM counterterms needed to renormalize them are then the following:
\begin{align}
 &-\tG^{(1)}_{\h} \underset{g_6\ \mathrm{UV}\ \mathrm{div.}}{=} \frac{3 M^2\Lambda}{8 \pi^2} g_6  \frac{1}{\epsilon} , \nonumber \\
 &-\tG^{(1)}_{\h\h} \underset{g_6\ \mathrm{UV}\ \mathrm{div.}}{=} -\frac{9 \Lambda^2}{4\pi^2} g_6^2 \frac{1}{\epsilon}+
\frac{3\Lambda }{8 \pi^2v}g_6 \left( 2 \MW^2 + \MZ^2+ 12 M^2 \right) \frac{1}{\epsilon} , \nonumber \\
&-\tG^{(1)}_{\h\h\h} \underset{g_6\ \mathrm{UV}\ \mathrm{div.}}{=} -\frac{81 \Lambda^2}{2\pi^2 v}g_6^2\frac{1}{\epsilon}+ \frac{9\Lambda }{8 \pi^2 v^2 }g_6 \left(2 \MW^2+ \MZ^2+ 29 M^2 \right)\frac{1}{\epsilon}, \nonumber\\
&-\tG^{(1)}_{\h^4} \underset{g_6\ \mathrm{UV}\ \mathrm{div.}}{=}-\frac{1593\Lambda^2}{4\pi^2v^2}g^2_6\frac1\epsilon-\frac{9\Lambda}{8\pi^2v^3}g_6\left(6\MW^2+3\MZ^2-119M^2\right)\frac1\epsilon,\nonumber\\
&-\tG^{(1)}_{\h^5} \underset{g_6\ \mathrm{UV}\ \mathrm{div.}}{=}-\frac{9585\Lambda^2}{4\pi^2v^3}g_6^2\frac1\epsilon-\frac{45\Lambda}{4\pi^2v^4}g_6\left(4\MW^2+2\MZ^2-27M^2\right)\frac1\epsilon,\nonumber \\
&-\tG^{(1)}_{\h^6} \underset{g_6\ \mathrm{UV}\ \mathrm{div.}}{=}-\frac{36045\Lambda^2}{4\pi^2v^4}g_6^2\frac1\epsilon-\frac{135\Lambda}{4\pi^2v^5}g_6\left(2\MW^2+\MZ^2-9M^2\right)\frac1\epsilon,\nonumber\\
&-\tG^{(1)}_{\h^7} \underset{g_6\ \mathrm{UV}\ \mathrm{div.}}{=}-\frac{19845\Lambda^2}{\pi^2v^5}g^2_6\frac1\epsilon,
\nonumber \\
&-\tG^{(1)}_{\h^8} \underset{g_6\ \mathrm{UV}\ \mathrm{div.}}{=}-\frac{19845\Lambda^2}{\pi^2v^6}g^2_6\frac1\epsilon.
\end{align}

%

\end{document}